\documentclass[onecolumn,noshowpacs,aps,prl,superscriptaddress]{revtex4}

\usepackage{graphicx}
\usepackage{dcolumn}
\usepackage{amsmath}
\usepackage{epsfig}

\input{symbols}

\newcommand{\BABARPubYear}    {10}

\newcommand{\BABARConfNumber} {003}
\newcommand{\SLACPubNumber} {14210}
\newcommand{\LANLNumber} {0000}

\begin{document}

\preprint{\babar-PUB-\BABARPubYear/\BABARConfNumber} 
\preprint{SLAC-PUB-\SLACPubNumber} 
\preprint{hep-ex\LANLNumber}

\begin{flushleft}
\babar -CONF-\BABARPubYear/\BABARConfNumber\\
SLAC-PUB-\SLACPubNumber\\
\end{flushleft}

\title{
{\Large \bf{Evidence for \boldmath{\btn} decays using hadronic $B$ tags}}
}

%
\author{P.~del~Amo~Sanchez}
\author{J.~P.~Lees}
\author{V.~Poireau}
\author{E.~Prencipe}
\author{V.~Tisserand}
\affiliation{Laboratoire d'Annecy-le-Vieux de Physique des Particules (LAPP), Universit\'e de Savoie, CNRS/IN2P3,  F-74941 Annecy-Le-Vieux, France}
\author{J.~Garra~Tico}
\author{E.~Grauges}
\affiliation{Universitat de Barcelona, Facultat de Fisica, Departament ECM, E-08028 Barcelona, Spain }
\author{M.~Martinelli$^{ab}$}
\author{D.~A.~Milanes}
\author{A.~Palano$^{ab}$ }
\author{M.~Pappagallo$^{ab}$ }
\affiliation{INFN Sezione di Bari$^{a}$; Dipartimento di Fisica, Universit\`a di Bari$^{b}$, I-70126 Bari, Italy }
% ...
\collaboration{The \babar\ Collaboration}
\noaffiliation

\date{\today}

\begin{abstract}

\noindent We present a search for the decay \btn\ using $467.8 \times 10^6$ \BB pairs   
collected at the \FourS resonance with the \babar\ detector 
at the SLAC PEP-II $B$-Factory. 
We select a sample of events with one completely reconstructed \Bm in an hadronic
decay mode ($B^-\to D^{(*)0}X^-$ and $\Bm \to \jpsi X^-$).
We examine the rest of the event to search for a \bptaunu decay. 
We identify the \taup lepton  in the following modes: $\tautoenunu$, $\tautomununu$,
$\tautopinu$ and $\tau^+ \to \rho \overline{\nu}_{\tau}$.
We find an excess of events with respect to expected background, which
excludes the null signal hypothesis at the level of 3.3 $\sigma$ and
can be converted to a branching fraction 
central value of  $\mathcal{B}(\btn)=( 1.80^{+0.57}_{-0.54}(\mbox{stat.}) \pm 0.26 (\mbox{syst.})) \times 10^{-4}$. 
\end{abstract}

\maketitle
\newpage

\section{\label{sec:level1}Introduction}
The study of the purely leptonic decay is 
of particular interest as a test of the Standard Model (SM) and a search 
for physics beyond the SM.
It is sensitive to the product of the $B$ meson decay constant $f_{B}$, 
and the absolute value of the Cabibbo-Kobayashi-Maskawa matrix element $\Vub$~\cite{ckm}.
In the SM the branching fraction is given by:
\begin{equation}
\label{eqn:br}
\mathcal{B}(B^{+} \rightarrow {\taup} \nu)= 
f_{B}^{2} \Vub^{2} \frac{G_{F}^{2} m^{}_{B}  m_{\tau}^{2}}{8\pi}
\left[1 - \frac{m_{\tau}^{2}}{m_{B}^{2}}\right]^{2} 
\tau_{\Bu}, 
\end{equation}
where  
$G_F$ is the Fermi constant,
$\tau_{\Bu}$ is the $\Bu$ lifetime, and
$m^{}_{B}$ and $m_{\tau}$ are the $\Bu$ meson and $\tau$ lepton masses.

The process is sensitive to possible extensions of the SM. 
For instance, in two-Higgs doublet models~\cite{twohiggs} and in minimal supersymmetric extensions 
of the SM it can be mediated by a charged Higgs boson.
A branching fraction measurement can therefore also be used to constrain the parameter space of extensions of the SM.
In a previously published analysis, based on a tagging technique using
hadronic $B$ decays that is similar to that used in this paper and a smaller data set, the \babar\ collaboration measured 
$\mathcal{B}(\btn) = (1.8^{+ 0.9}_{ -0.8} \mbox{(stat.)} \pm 0.4 \pm 0.2 \mbox{(syst.)} )\times 10^{-4}$~\cite{babarhad0}, 
and using tagging based on reconstruction of semileptonic $B$ decays 
$\mathcal{B}(\btn) = (1.7 \pm 0.8 \mbox{(stat.)} \pm 0.2 \mbox{(syst.)} )\times 10^{-4}$ 
~\cite{babarsl}.
The Belle collaboration measured, with a similar tagging technique used in this analysis, the branching fraction to be
$\mathcal{B}(\btn) = (1.79 ^{+ 0.56}_{ -0.49}  \mbox{(stat.)} ^{+ 0.46}_{ -0.51} \mbox{(syst.)}  ) \times 10^{-4}$~\cite{bellehad},
and using a tagging algorithm based on the reconstruction of semileptonic $B$ decays 
$\mathcal{B}(\btn) = (1.54 ^{+ 0.38}_{ -0.37} \mbox{(stat.)} ^{+ 0.29}_{ -0.31} \mbox{(syst.)} ) \times 10^{-4}$~\cite{bellesl}.

\section{The \babar\ detector and dataset}
\label{sec:babar}

The data used in this analysis were collected with the \babar\ detector
at the \pep2\ storage ring. 
The sample corresponds to an integrated
luminosity of \onlumi at the \FourS\ resonance (on-resonance) 
and \offlumi taken at $40\mev$ below the $B\bar{B}$ production threshold 
(off-resonance), which is used to study background from
$e^+e^-\to f\bar{f}$ ($f = u, d, s, c, \tau$) continuum events. 
The on-resonance sample contains $(467.8 \pm 5.1) \times 10^{6}$ \BB decays. 
The detector is described in detail elsewhere~\cite{nimbabar}.
Charged particle trajectories are measured in the tracking system
composed of a five-layer silicon vertex detector and a 40-layer drift chamber (DCH),
operating in a  1.5~T solenoidal magnetic field.
A Cherenkov detector is used for charged $\pi$--$K$ discrimination, a CsI calorimeter (EMC)
for photon and electron identification, and 
the flux return of the solenoid, which consists of layers of iron 
interspersed with resistive plate chambers or limited streamer tubes, for muon
and neutral hadron identification. 

In order to estimate signal selection efficiencies and to study physics backgrounds,
we use a Monte Carlo (MC) simulation based on \geantfour~\cite{geant4}.
In MC simulated signal events one \Bp meson decays as \btn  and the other decays in any final state.
The \BB\ and continuum MC samples are equivalent to approximatively three times and
1.5 times, respectively, the accumulated data sample.
Beam-related background and detector noise are taken from data 
and overlaid on the simulated events.

\section{Signal selection}
\label{sec:selection}
We reconstruct an exclusive decay of one of the \B mesons in the event
(which we refer to as the tag $B$) and examine the rest of the event
for the experimental signature of \btn (charged-conjugate modes are implied throughout the paper).
We consider the most abundant $\tau$ decay modes \tauex, \taumux, \taupix, \taurhox, totaling approximatively 70\% of all $\tau$ decays.
The signal region in data is kept blind until the end of the analysis chain when we extract the signal yield.
We reconstruct the tag \B candidate in the set of hadronic decays $B^-\to M^0 X^-$, where $M^0$ denotes a $D^{(*)0}$ or a  \jpsi, and 
$X^-$ denotes a system of hadrons with total charge $-1$ composed of
$n_1 \pi^{\pm}$, $n_2 K^{\pm}$, $n_3 \pi^0$, $n_4 \KS$ where
$n_1 + n_2 \le 5$, $n_2 \le 2$, $n_3$~and~$n_4 \le 2$.
We reconstruct the $D^0$ as $D^0  \to K^- \pi^+, 
K^- \pi^+ \pi^0, K^- \pi^+ \pi^- \pi^+, \KS \pi^0, \KS \pi^+ \pi^-,  \KS \pi^+ \pi^- \pi^0, K^+ K^-, \pi^+ \pi^-$.  
We reconstruct the \Dstarz meson as $\Dstarz \to D^0 \pi^0, D^0 \gamma$, and the \jpsi meson as $\jpsi \to e^+e^-, \mu^+\mu^-$.
The kinematic consistency of the tag \B candidates is checked with the
beam energy-substituted mass $\mes = \sqrt{s/4 - p_B^2}$ and the
energy difference $\DeltaE=E_B-\sqrt{s}/2$, where $\sqrt{s}$ is the
total energy in the $\Upsilon(4S)$ center of mass system and $p_B$ and
$E_B$ denote respectively the momentum and the energy of the tag \B
candidate in the center of mass frame. The resolution on \DeltaE is
measured to be $\sigma_{\DeltaE} = 10-35 \mev$, depending on the decay
mode; we require $|\DeltaE| < 3\sigma_{\DeltaE}$. 
Events with a candidate tag \B arise from two possible classes with different \mes distributions.
Signal events with a correctly reconstructed tag \B and the other \B decaying as \btaunu,
and background events from $\FourS \to \BpBm $ with a correctly
reconstructed tag \B  are characterized by an \mes distribution peaked 
at the \B mass. The other class of events consists of continuum background, $\epem \to \q \qbar $ (\q = \u, \d, \s, \c) and $\epem \to \tautau$, 
and combinatorial background, $\Y4S \to \BzBzb$ or \BpBm in which the tag \B is misreconstructed; this class of events has a broad \mes distribution that can be modeled by means of a phenomenological threshold function (ARGUS function)~\cite{argus}.
 
\begin{figure}[!tbh]
\begin{center}
\begin{tabular}{c}
\includegraphics[width=0.6\textwidth]{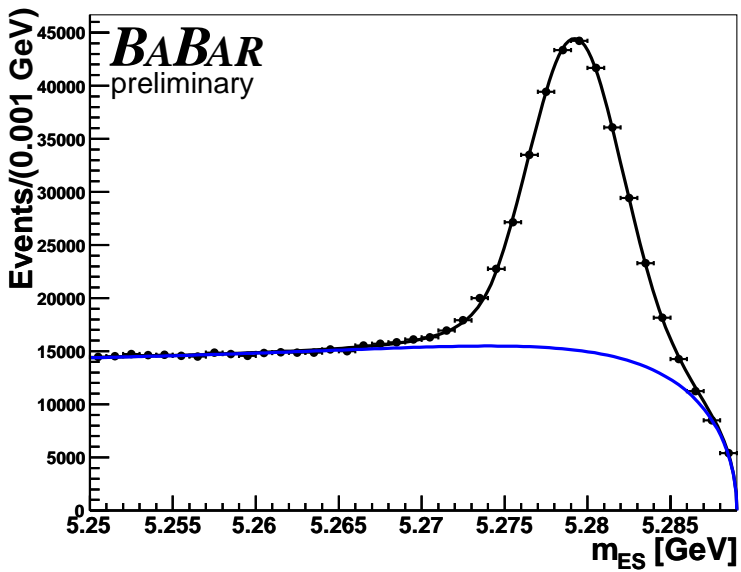}
\end{tabular}
\caption{Fit to the \mes distribution in data. Dots are data, the blue curve represents the fitted combinatorial and continuum background.}
\label{fig:mes_data}
\end{center}
\end{figure}

If multiple tag \B candidates are reconstructed we select that with the lowest value of $|\Delta E|$.
The purity ${\cal P}$ of each reconstructed \B decay mode is estimated as the ratio of the number of peaking events with $m_{ES} > 5.27 \gev$ to the total number
of events in the same range. We consider only events with the tag \B reconstructed in decay modes with ${\cal P} > 0.1$.
The yield in data is determined by means of an extended unbinned maximum likelihood fit to the \mes distribution, as shown in figure~\ref{fig:mes_data}.
We use as probability density function (PDF) for the combinatorial and continuum background
an ARGUS function, while for the correctly reconstructed tag \B component we use as PDF a
Gaussian function with an exponential tail (Crystal Ball function)~\cite{cball}.
Combinatorial and continuum backgrounds in any discriminating variable are estimated 
from a sideband in \mes ($ 5.209 \gev < \mes < 5.260 \gev$) and extrapolated into the signal region ($\mes > 5.270 \gev$) using the
results of a fit to an ARGUS function.
The peaking \BpBm background is determined from \BpBm MC, after subtraction of
the combinatorial component to avoid double counting by means of a
similar fit.
 
After the reconstruction of the tag $B$,  we apply a set of selection criteria on the rest of the event.
We require the presence of only one well reconstructed charged track (signal track), with charge opposite to that of the tag $B$. 
The $\tau$ lepton is reconstructed in one of four decay modes: \tauex, \taumux, \taupix, \taurhox.
We separate the event sample in four categories using particle identification 
criteria applied to the signal track. The \taurho sample is obtained by associating the signal
track with a $\pi^0$ reconstructed from a pair of neutral clusters with invariant mass
between 115 \mevcc and 155 \mevcc.   
In order to remove the $\epem \to \tautau$ background we impose \mtau mode dependent
requirements, preserving 90\% of the \btn signal, on the ratio between the $2^{nd}$ and the $0^{th}$ Fox-Wolfram moments (R2)~\cite{r2}
calculated using all the charged tracks and neutral clusters of the event.

In order to reject the continuum and combinatorial background we use discriminating variables 
constructed from the kinematics of the tag \B candidate.
The first variable is the momentum in the CM frame ($p^*_M$) of the $D^{(*)0}$ or \jpsi candidate reconstructed from
the decay products of the tag $B$. 
The second variable is the absolute value of the thrust~\cite{thrust} ($|\vec{T_B}|$) of the tag $B$.
The third variable is the cosine of the angle between
the thrust of the tag \B and the thrust of the rest of the event ($\cos \theta_{TB}$). 
We combine $p^*_M$,  $|\vec{T_B}|$ and $\cos \theta_{TB}$ in a likelihood ratio 
$L_C = L_S( p^*_M, |\vec{T_B}|, \cos \theta_{TB} ) /   (L_S( p^*_M, |\vec{T_B}|, \cos \theta_{TB} ) +  L_B( p^*_M, |\vec{T_B}|, \cos \theta_{TB} ))$,
where the signal ($S$) and background ($B$) likelihoods are obtained
from the products of the PDFs of the three discriminating variables:
$L_{S}( p^*_M, |\vec{T_B}|, \cos \theta_{TB}  ) =  P_{S}( p^*_M
) P_{S}( |\vec{T_B}| ) P_{S}( \cos \theta_{TB}  )$ and 
$L_{B}( p^*_M, |\vec{T_B}|, \cos \theta_{TB}  ) =  P_{B}( p^*_M
) P_{B}( |\vec{T_B}| ) P_{B}( \cos \theta_{TB}  )$.
The PDFs for the signal modes are obtained from the signal MC, 
whereas the PDFs for backgrounds are obtained from the \mes sideband
in data.
 
In order to further reject the background from correctly reconstructed tag \B events, we 
impose a requirement on center of mass momentum of the signal track for the \taue, \taumu and \taupi  modes.
For the \taurho mode we combine in a likelihood ratio ($L_P$) the following variables:
the invariant mass of the signal track and the \piz,  the total momentum in the CM frame of the
pair $|\vec{p}^*_{\rho}|$, the momentum in the CM frame of the \piz, and the missing mass of the event.
The PDFs used in the likelihood ratio for the signal and background are determined from signal and \BpBm MC, respectively.

The most discriminating variable is \eextra, defined as the sum of the energies of
the neutral clusters not associated with the tag \B or with the signal $\pi^0$
from the \taurho mode, and passing a minimum energy requirement (60 \mev).
Signal events tend to peak at low \eextra, background events, 
which contain additional sources of neutral clusters, tend to be distributed at higher values.

We optimize the selection requirements, including those on the purity ${\cal P}$ of the tag \B and
the minimum energy of the neutral clusters, aiming at the lowest expected uncertainty in 
the branching fraction fit. In order to estimate the uncertainty, which includes
the statistical and the largest systematic uncertainties, we run 1000 toy experiments
extracted from the background and signal expected shapes for a set of possible selection requirements,
assuming a branching fraction of $1.4 \times 10^{-4}$\cite{pdg}.

\begin{figure}[!tbh]
\begin{center}
\begin{tabular}{cc}
\multicolumn{2}{c}{\includegraphics[width=0.8\textwidth]{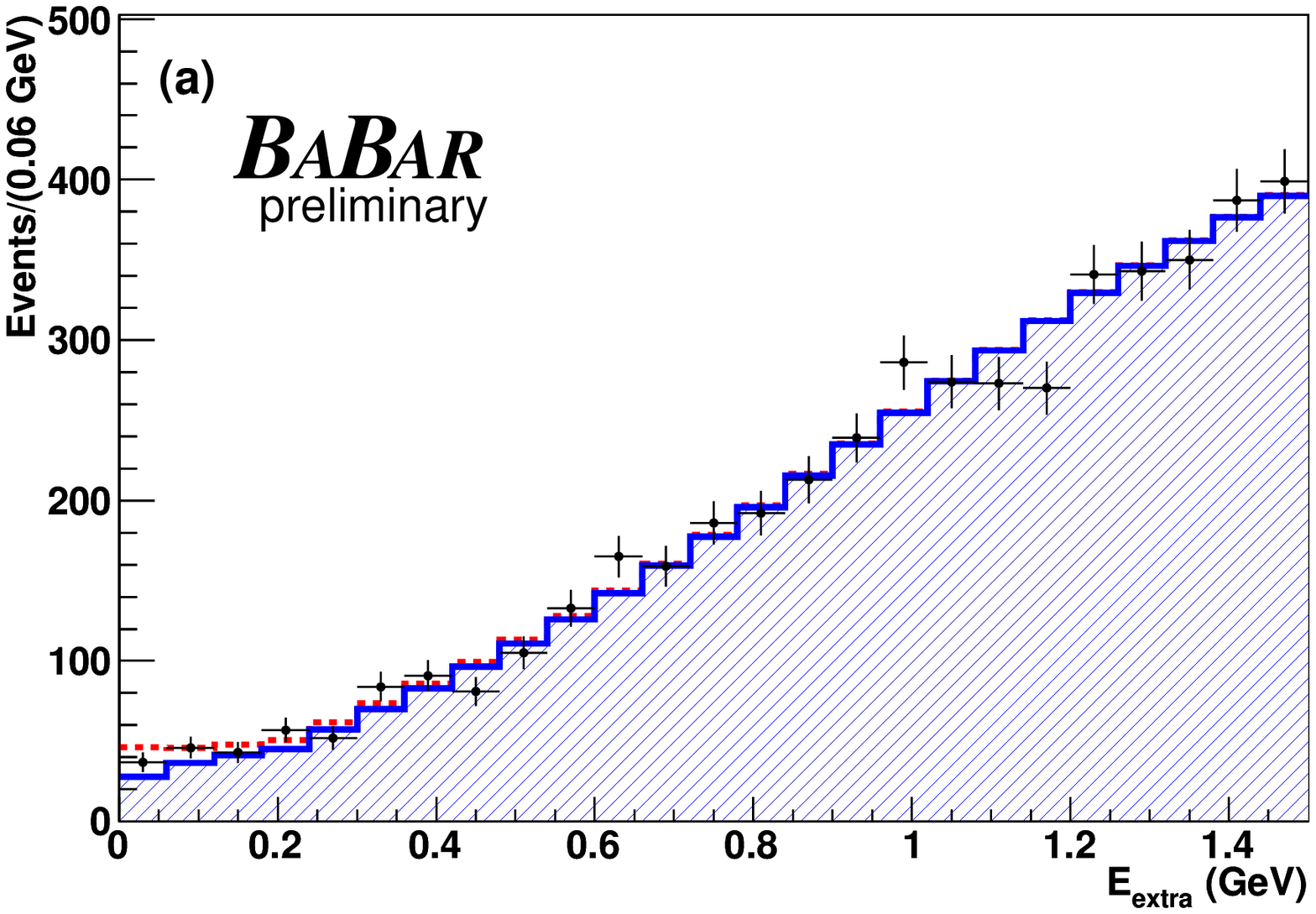}} \\
\includegraphics[width=0.4\textwidth]{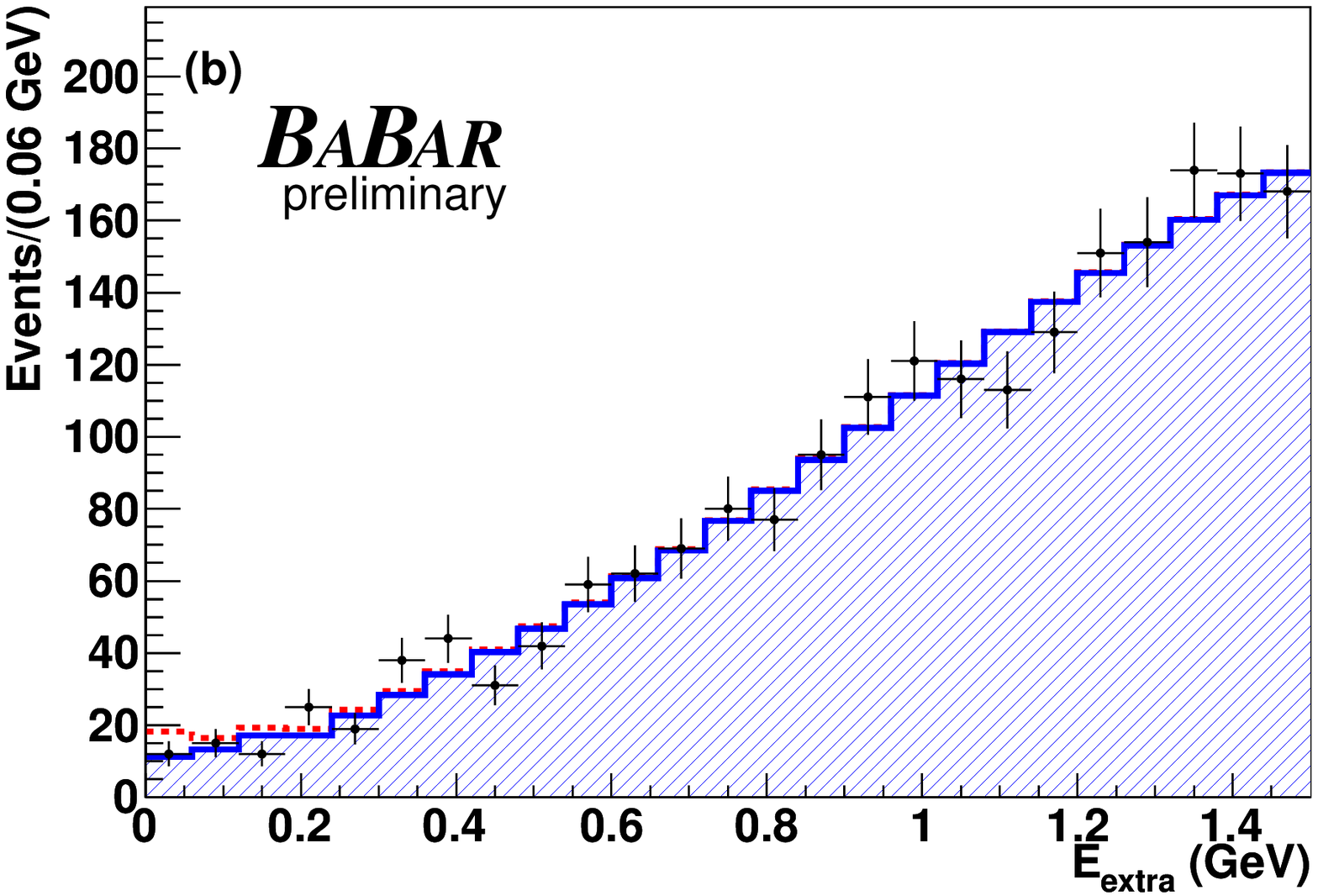} &
\includegraphics[width=0.4\textwidth]{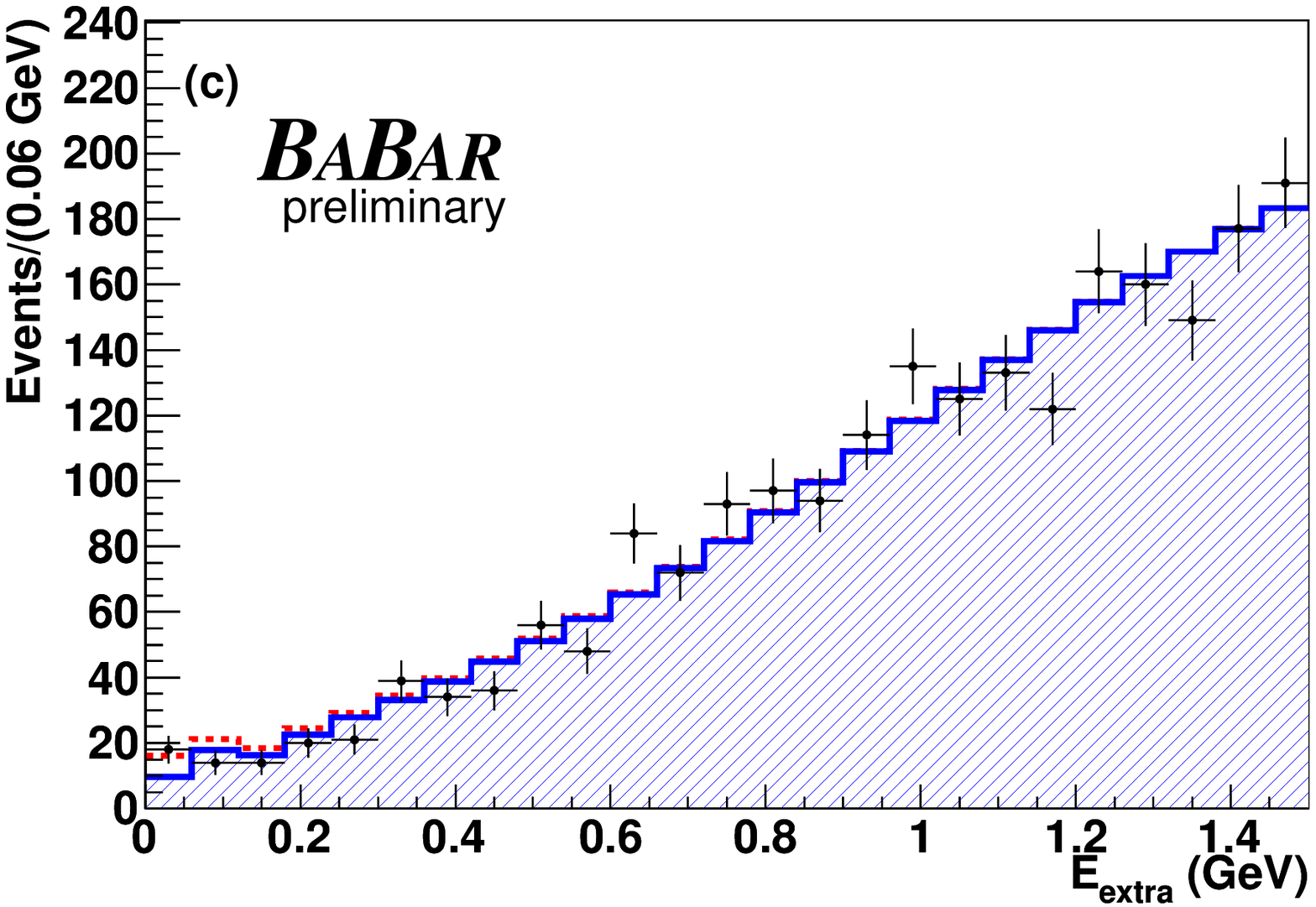} \\
\includegraphics[width=0.4\textwidth]{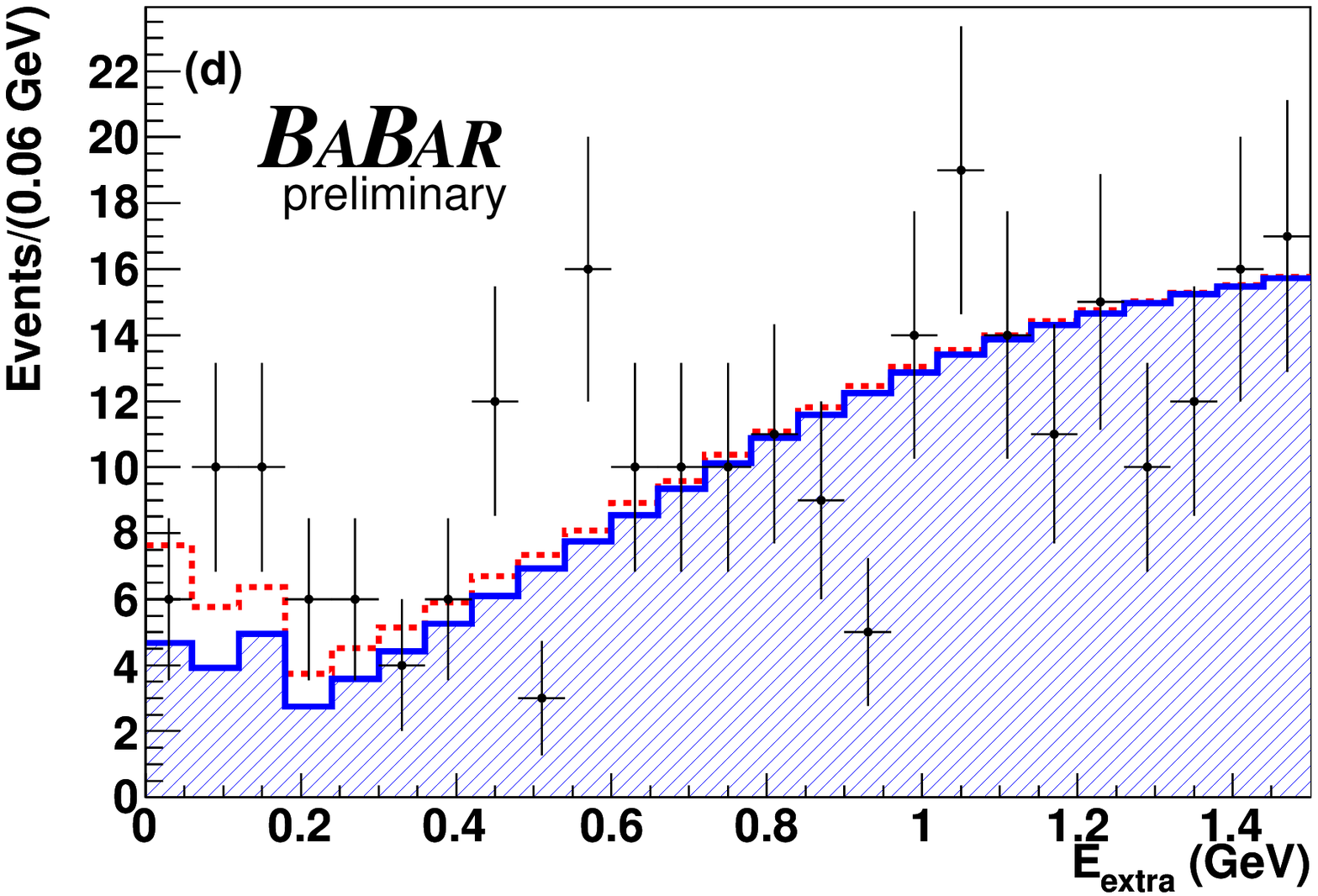} &
\includegraphics[width=0.4\textwidth]{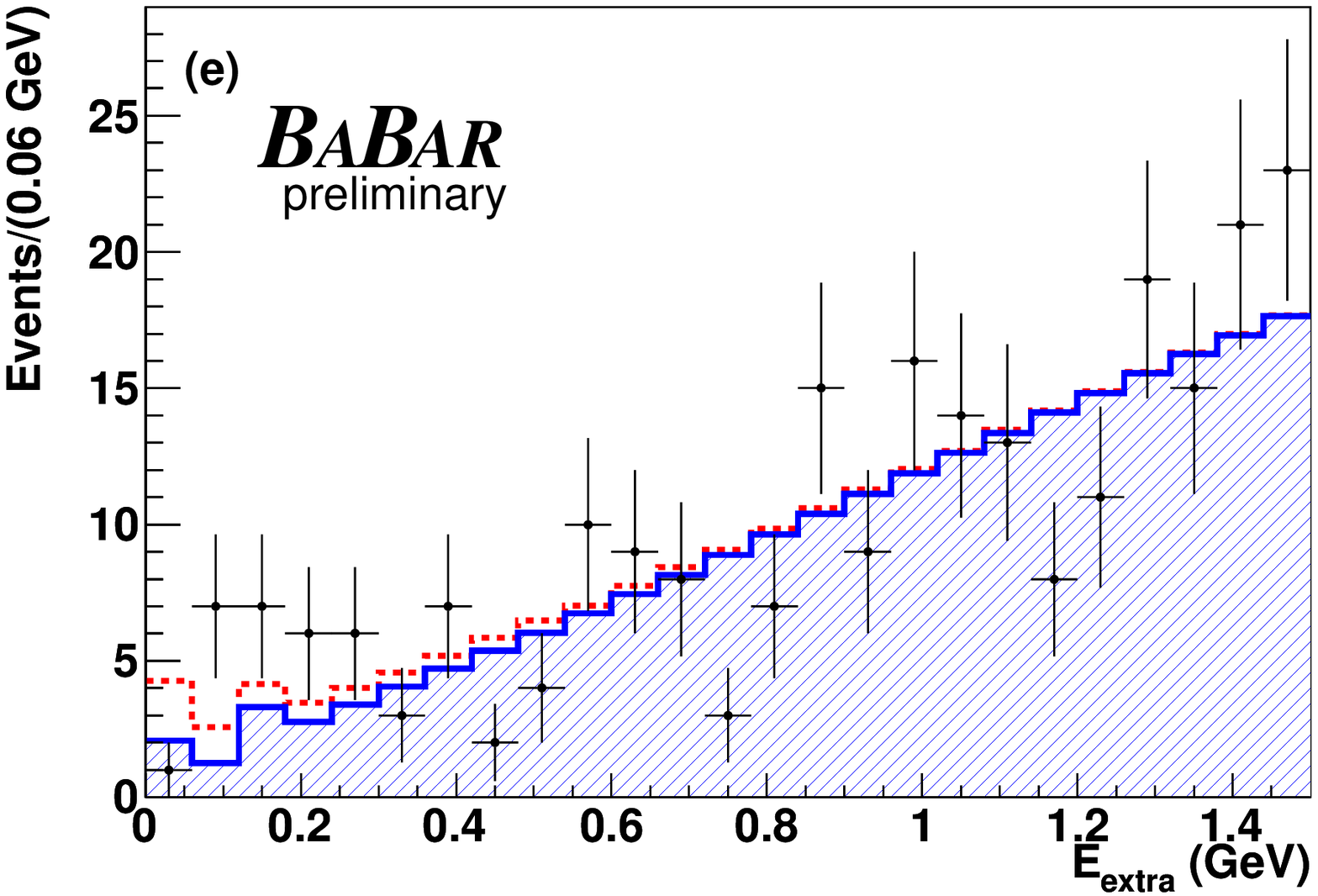} 
\end{tabular}
\caption{\eextra distribution in data (dots with error bars) with all selection
 requirements applied and fit results overlaid. The hatched histogram is
 the background, the red dashed component is the signal. 
Plot \emph{(a)} shows all $\tau$ decay modes fitted simultaneously. Lower plots show the projection of the
simultaneous fit result on the four analyzed $\tau$ decay modes: \emph{(b)} \taue, \emph{(c)} \taumu,
\emph{(d)} \taupi, \emph{(e)} \taurho.}
\label{fig:eextrasel}
\end{center}
\end{figure}

The signal selection requirements are summarized in table \ref{tab:selectiontable}.  
The \eextra distribution with all the selection requirements applied is shown in  figure \ref{fig:eextrasel}.

\begin{table}[!tbh]
\begin{center}
\begin{tabular}{lcccc}
\hline
\hline 
 Variable &\taue&\taumu&\taupi&\taurho\\
\hline
purity & \multicolumn{4}{c}{$>10\%$} \\ 
cluster energy (\mev)&  \multicolumn{4}{c}{60}  \\
R2  &$<0.57$&$<0.56$&$<0.56$&$<0.51$\\
$L_C$ &$>0.2$ &  $> 0$& $>0.3$&$>0.45$\\
$p^*_{trk} (\gevc)$& $<2.1$&$<2 $& $>1.4$ & \\
$L_P$ & & & &$>0.8$\\
\hline
\hline
\end{tabular}
\caption{Optimized signal selection criteria for each \mtau mode.}
\label{tab:selectiontable}
\end{center}
\end{table}

\section{Branching Fraction measurement procedure and results}
We use an extended unbinned maximum likelihood fit to extract
the $B^+\to\tau^+\nu$ branching fraction.  
The likelihood function for the $N_k$ candidates  reconstructed in one of the four $\tau$ decay modes $k$ is
\begin{equation}
\label{eq:pdfsum}
{\cal L}_k = e^{-(n_{s,k}+n_{b,k}) }\!\prod_{i=1}^{N_k}
                \bigg\{ n_{s,k} 
         {\cal P}_{k}^{s}(E_{i,k})
        + n_{b,k} {\cal P}_{k}^{b}(E_{i,k}) 
        \bigg\}
\end{equation}
where $n_{s,k}$ is the signal yield, $n_{b,k}$ is
the background yield,
$E_{i,k}$ is the \eextra value of the $i^{th}$ event, ${\cal P}_k^s$ is the
probability density function of signal events, and ${\cal P}_k^b$ is the
probability density function of background events.
The background yields in each decay mode are permitted to float independently of each other in the fit, while the 
signal yields are constrained to a single branching ratio via the relation:
\begin{equation}
\label{eq:BFcalc}
 n_{s,k} = N_{ B\overline{B}} \times \epsilon_k\times {\cal B}
\end{equation}
where $N_{ B\overline{B}} = (4.678  \pm 0.051) \times 10^{8}$ is the number of $B\overline{B}$ pairs in the data sample, $\epsilon_k$
is the $\tau$ decay mode dependent reconstruction efficiency, and
${\cal B}$ is the $B^+\to\tau^+\nu$ branching fraction. 
The parameters $N_{ B\overline{B}}$ and $\epsilon_k$  are fixed in the
fit while ${\cal B}$ is left floating. 
The reconstruction efficiencies $\epsilon_k$, which include the $\tau$ branching fractions, are obtained from MC simulation of the signal.
Since the tag \B reconstruction efficiency is included in $\epsilon_k$ and is estimated from the signal MC, we apply a correction factor
$R_{\rm{data/MC}} = 0.926 \pm 0.010$
to take into account data/MC differences,  taking the
ratio of the peaking component of the \mES\ distribution of the hadronic tag \B in data and in MC simulation events.

\begin{figure}[!tbh]
\begin{center}
\begin{tabular}{c}
\includegraphics[width=0.8\textwidth]{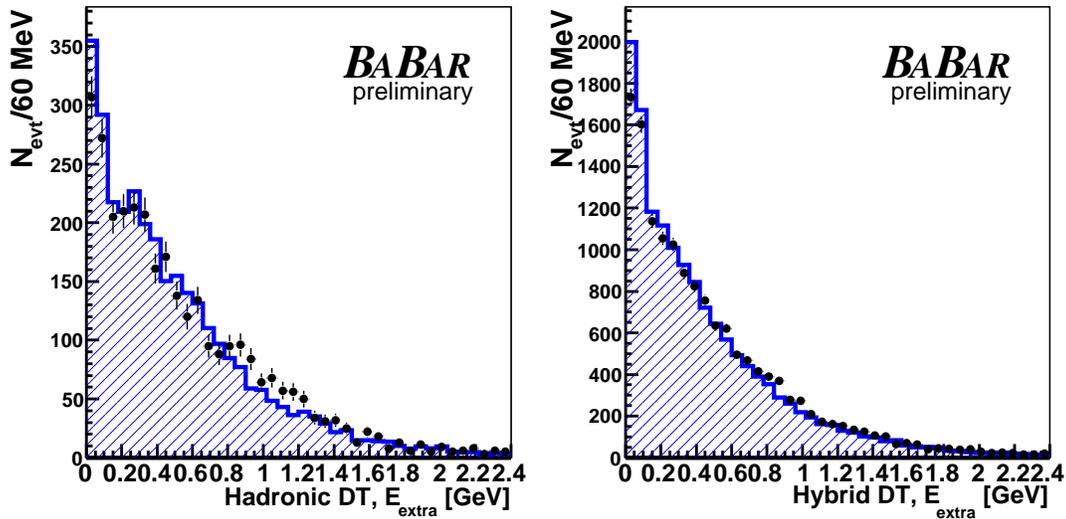}
\end{tabular}
\caption{\eextra distribution for double tags. The second \B is reconstructed in hadronic decays (left plot) or semileptonic decays (right plot). Points are data, histograms are MC simulation.}
\label{fig:doubletags}
\end{center}
\end{figure}

We use histograms with a bin width of 60 \mev to represent the PDFs  ${\cal P}_{k}^{s}$ and ${\cal P}_{k}^{b}$  for signal and background, respectively.
The signal PDF is obtained from a high statistics signal MC
simulation sample, corrected for data/MC disagreement.
Since a data sample of suitable statistics with exactly the same final states as our signal channels 
is not available, we use a sample of fully reconstructed events where in addition to the reconstructed tag $B$, a second
\B is reconstructed in an hadronic or a semileptonic decay mode, using
charged tracks and neutral clusters not assigned to the tag $B$.
In order to estimate the correction to the signal PDF, we compare the
distribution of \eextra in this double tags sample from
experimental data and MC simulation. 
The distributions are normalized to the same area and the comparison is shown in figure \ref{fig:doubletags}.
We extract the correction function taking the ratio of the two
distributions and fitting it with a second order polynomial. 

We take the PDF of the combinatorial background from the \mes sideband. 
The contribution of this component  in the signal region is  obtained by fitting the \mes~distribution after the selection has been applied.
The shape of the peaking background is taken from \BpBm MC in the signal region, after the intrinsic combinatoric
background has been subtracted by a fit to \mes, to avoid double counting.
The two background components  are added together in a single background PDF. 
We finally apply a smoothing procedure on the total background shape,
excluding the first bins.
We estimate the branching fraction minimizing $-2 \ln {\cal L}$, where
${\cal L}  = \Pi_{k=1}^{4} {\cal L}_k$, and
${\cal L}_k$ is defined in equation~\ref{eq:pdfsum}.The projections of the fit results are shown in figure \ref{fig:eextrasel}.
We observe a significant excess of events with respect to the expected
backgrounds and measure a branching fraction 
${\cal B}( \btn) = ( 1.80 ^{+0.57}_{-0.54} )\times 10^{-4}$, where the
uncertainty is statistical. 
We evaluate the significance of the observed signal, including only
statistical uncertainty, as $S = \sqrt{2 \ln({\cal L}_{s+b}/{\cal L}_{b})}$, where
${\cal L}_{s+b}$ and ${\cal L}_{s}$ denotes the obtained maximum likelihood value and the likelihood value
assuming background only. We find $S = 3.6\sigma$. Table \ref{tab:res} summarizes the results from the fit.

\begin{table}
\begin{center}
\begin{tabular}{lccc}
\hline
\hline
Decay Mode    &  $\epsilon \times 10^{-4}$& Branching Fraction ($\times 10^{-4})$ & Significance $\sigma$ \\
\hline
\taue  &   $2.73$     & $0.39 ^{+0.89}_{-0.79}$  &  0.5 \\
\taumu &  $2.92$    & $1.23 ^{+0.89}_{-0.80}$  & 1.6  \\
\taupi &  $1.55$      & $4.0   ^{+1.5}_{-1.3}$   & 3.3  \\
\taurho & $0.85$     & $4.3   ^{+2.2}_{-1.9}$   & 2.6  \\
\hline
combined & $8.05$  & $1.80^{+0.57}_{-0.54}$  & 3.6 \\
\hline
\hline
\end{tabular}
\caption{Reconstruction efficiency $\epsilon$, measured branching fractions and statistical significance
obtained from the fit with all the modes separately and constrained to
the same branching fraction. The $\tau$ decay mode branching fractions are included in the efficiencies.}
\label{tab:res}
\end{center}
\end{table} 
\section{Systematics}

The dominant source of systematic uncertainty is the background
PDF, due the finite statistics of the \BpBm MC simulated sample, used
to estimate the \BpBm background PDF, and of the \mes data sideband, used
to estimate the combinatorial and continuum backgrounds. In order to 
estimate the systematic uncertainty we repeat the fit of the branching
fraction with 1000 variations of the background PDF, varying each bin
within the statistical error, and assign 12\% as systematic uncertainty. 

The systematic uncertainty
due to the signal \eextra distribution correction function
obtained from data/MC comparisons using control samples is obtained by
varying the parameters of the second order polynomial within their uncertainty and repeating
the fit to the \btn branching fraction. We observe a 1.7\% 
variation that we take as the systematic uncertainty on the signal shape. 
 
Uncertainty in the differences between data and MC in the tracking and neutral
reconstruction efficiencies reflects in the uncertainty in the central value
of the branching fraction. The difference of the tracking efficiency is estimated
with a control sample of high momentum tracks from $\epem \to \tautau$ events
to be 0.5\% per track. Since there is only one signal track candidate in all four 
$\tau$ decay modes in the \btn signal, we use this value as the
uncertainty due to tracking efficiency. We accept events with one extra low $p_{T}$
charged track.
Comparing the multiplicity of low $p_T$ charged tracks from the double tags sample 
in data and in MC,  we estimate the systematic uncertainty 
to be 1.3\%.  Adding in quadrature the two uncertainties we estimate the the systematic
error to be 1.4\%.

Other systematic uncertainties on the efficiency stem from the finite signal MC statistics (0.8\%),
the uncertainty in the tag \B efficiency correction (5.0\%), the
electron identification (2.6\%), and muon identification (4.7\%). 
The systematic uncertainties are summarized in Table \ref{tab:systematics}.
The total systematic uncertainty is obtained by
combining all sources in quadrature.

\begin{table}
\begin{center}
\begin{tabular}{lc}
\hline
\hline
Source of systematics   & BF uncertainty (\%)\\
\hline
\B counting                      &  0.5\\
Tag \B efficiency            &   5.0 \\  
Background PDF            &  12    \\
Signal PDF                     &  1.7   \\
MC statistics                 &  0.8        \\
Electron identification   &  2.6  \\  
Muon identification       &  4.7        \\  
Kaon identification        &   0.4\\
Tracking                        &   1.4  \\
\hline
Total                               &  14  \\
\hline
\hline
\end{tabular}
\caption{Contributions to systematic uncertainty on the branching
 fraction.}
\label{tab:systematics}
\end{center}
\end{table}
\section{Conclusions}
In summary, we have measured the branching fraction of the decay \btn
using a tagging algorithm based on the reconstruction of hadronic \B
decays using a data sample containg $467.8 \times 10^6$ \BB pairs collected
with the \babar\ detector at the \pep2 $B$-Factory.
We measure the branching fraction to be 
$\mathcal{B}(\btn)=( 1.80^{+0.57}_{-0.54}(\mbox{stat.}) \pm 0.26 (\mbox{syst.})) \times 10^{-4}$,
excluding the null hypothesis at the level of 3.6 standard
deviations using statistical uncertainties only, and at the level of
3.3 standard deviations including the systematic uncertainties.
 This result supersedes our previous result using the same technique~\cite{babarhad0}.
Combining this result with the other \babar\ measurement of
$\mathcal{B}(\btn)$  derived from a statistically
independent sample~\cite{babarsl}, we obtain a single \babar\ result 
$\mathcal{B}(\btn)=( 1.76 \pm 0.49 ) \times 10^{-4}$, where the
uncertainty includes both statistical and systematic uncertainties.

\section{Acknowledgements}
We are grateful for the 
extraordinary contributions of our \pep2\ colleagues in
achieving the excellent luminosity and machine conditions
that have made this work possible.
The success of this project also relies critically on the 
expertise and dedication of the computing organizations that 
support \babar.
The collaborating institutions wish to thank 
SLAC for its support and the kind hospitality extended to them. 
This work is supported by the
US Department of Energy
and National Science Foundation, the
Natural Sciences and Engineering Research Council (Canada),
the Commissariat \`a l'Energie Atomique and
Institut National de Physique Nucl\'eaire et de Physique des Particules
(France), the
Bundesministerium f\"ur Bildung und Forschung and
Deutsche Forschungsgemeinschaft
(Germany), the
Istituto Nazionale di Fisica Nucleare (Italy),
the Foundation for Fundamental Research on Matter (The Netherlands),
the Research Council of Norway, the
Ministry of Education and Science of the Russian Federation, 
Ministerio de Ciencia e Innovaci\'on (Spain), and the
Science and Technology Facilities Council (United Kingdom).
Individuals have received support from 
the Marie-Curie IEF program (European Union), the A. P. Sloan Foundation (USA) 
and the Binational Science Foundation (USA-Israel).

\bibliographystyle{apsrev}
\end{document}